\documentclass[12pt]{iopart}
\usepackage[pdftex]{graphicx}
\begin{document}
\title[Fabrication of silicon nanocrystals ......]
{Fabrication of silicon nanocrystals using sequential Au ion implantation}
\author{Gayatri Sahu$^1$ \footnote{Electronic mail: gayatri@iiti.ac.in}, \ Rajesh\  Kumar$^{1,2}$\  and \    D. P.\  Mahapatra$^3$}
\address{$^1$ Discipline \  of \  Physics, \ School\  of \  Basic \ Sciences, \ Indian \ Institute \  of \ Technology \ Indore, \ Madhya \ Pradesh-452017, \ India}
\address{$^2$ Surface \  Science  \ and \ Engineering \  Group, \ Indian \ Institute \  of \ Technology \ Indore, \ Madhya \ Pradesh-452017, \ India }
\address{$^3$ Institue \ of\  Physics, \ Sachivalaya \ Marg, \ Bhubaneswar-751005, \ Odisha, \ India}  

\begin{abstract}

Silicon nanocrystals are produced using a two-stage gold ion implantation technique. First 
stage implantation using low energy ions leads to the formation of an amorphous Si (a-Si) 
layer. A subsequent high energy Au irradiation in the second stage is found to produce 
strained Si NCs. An annealing at a temperature as low as 500$^o$C is seen to result in strain free NCs showing quantum confinement effects. Higher temperature annealing of the samples is found to result in growth in size from recrystallization of the a-Si matrix. Raman Scattering, X-ray diffraction (XRD) and Rutherford Backscattering spectrometry (RBS) have been used to study the effect of annealing on the samples and the size of Si NCs formed. The data could be well explained using a phonon confinement model with an extremely narrow size distribution. XRD results go in line with Raman analysis. 

\end{abstract}

\maketitle
\section{Introduction}

Silicon (Si) dominates the electronic industry but its poor optical properties due to indirect band gap, restrict it to be used in photonic applications. This can be circumvented in going to nano-meter length scales \cite{canham}. At sizes comparable to the excitonic Bohr radius ($a_B$ = 4.3 nm), Si nanocrystals (NCs) show quantum confinement effect. At these sizes, the carrier (electron or hole) wave-functions spread out in $k$-space, breaking the crystal momentum selection rule, leading to quasi-direct transitions \cite{hybertsen}. This leads to the possibility of achieving efficient light emission from Si nanostructures resulting in industrially viable monolithically integrated Si based optoelectronic devices and systems. Visible emissions from Si NCs in the range of red to blue have been observed depending on the particle size which has been attributed to quantum confinement effects. Surface oxidation also affects the visible photoluminescence (PL) from Si NCs \cite{wolkin,puzder}. To avoid such effects, embedded NCs in a given matrix are useful. For this, ion beam synthesis has proven to be a suitable method in producing embedded NCs. This method is a widely used technique in Si processing involving VLSI applications. The research activity in Si technology is stimulated by the possibility to produce high volume fraction of NCs with controllable size and potential applications in nano-electronics and nano-photonics \cite{gonella,meldrum,mazzoli}. In view of this, we have tried to synthesize Si NCs embedded in bulk Si using a two stage heavy ion implantation/irradiation technique \cite{nanoGS,ftcGS}. In this, a low energy (keV) Au implantation has been used to get a $\sim$ 30 nm thick a-Si layer. Si NCs are formed in this amorphised layer through an MeV Au ion induced localized crystallization. Annealing of the prepared samples also changes the Si NCs size. Since, these Si NCs are embedded in bulk Si, no  surface oxidation is expected giving an additional advantage. The initial low energy Au implantation for amorphisation also introduces a significant amount of Au which is also known to form NCs dispersed in the top a-Si matrix when annealed. This has been found to be very important regarding enhancement in luminescence of the Si NCs produced \cite{ftcGS}. At high fluence, the implanted gold atoms can also form localized gold-silicide (Au-Si) complexes \cite{nanoGS} even in as-implanted samples. Production of these together with Au particles inside and in the vicinity of the Si NCs is expected to produce some strain in them.

Raman Spectroscopy is a powerful tool for investigating of radiation-induced damage in Si. This is possible because the intensity, half-width and peak shift of the zone center phonon peak are very sensitive to structural disorder, damage and stress in the lattice. In addition to this, using the phonon confinement model, Raman Spectroscopy provides a tool to estimate the size of NCs in case they are present in the system \cite{ritcher}.  In the present case, the dependence of annealing temperature on the NC formation and size has been studied, using Raman scattering and XRD. Using an RBS analysis, we have investigated the annealing kinetics and diffusion of Au in Si. This analysis gives the information regarding Au distribution in Si matrix with annealing temperature and its effect on Si NC formation/size variation.

\section{Experiment}

In the present study, Si(100) samples (n-type, P doped with resistivity 1 - 20 $\Omega$-cm), were sequentially implanted/irradiated with 32 keV and 1.5 MeV Au ions with an ion fluence 
of $4 \times 10^{16}$ cm$^{-2}$ and $1 \times 10^{15}$ cm$^{-2}$, respectively. The implantations were carried out at beam currents $\sim$ 240 nA and 40-60 nA for low and high energy implantation cases, respectively. During implantation, samples were tilted 7$^o$ from the axis of ion beam for reducing any channeling effect. The fluence of keV Au implantation was chosen by the fact that above a fluence of $3 \times 10^{16} cm^{-2}$ there is no change in the optical properties due to sputtering effects \cite{vacuumGS}. The doubly Au implanted samples were annealed at five different temperatures varying between 500$^o$ - 950$^o$ C for 1 hr in air. The sample was introduced into the furnace directly at the specified temperature and annealed for 1 hr. After that, furnace power was switched off resulting in slow cooling upto room temperature. 

Raman scattering measurements were carried out in back-scattering geometry using a 
RAMANOR U1000 Jobin-Yvon micro-Raman spectrometer equipped with liquid nitrogen 
cooled charged-coupled device as the detector. An Ar-ion laser tuned to 514.5 nm was used 
as the excitation source. RBS measurements were carried out with 1.35 MeV He$^+$ ions at a backscattering angle of 165$^o$. One of the samples was subjected to high resolution transmission electron microscopy (TEM) for a direct look at the buried NCs. All the implantations and irradiations were carried out at room temperature. 

\section{Results and Discussion}

\subsection{Raman Scattering results :}

\begin{figure}[h]
\begin{center}
\includegraphics[height=4.0cm]{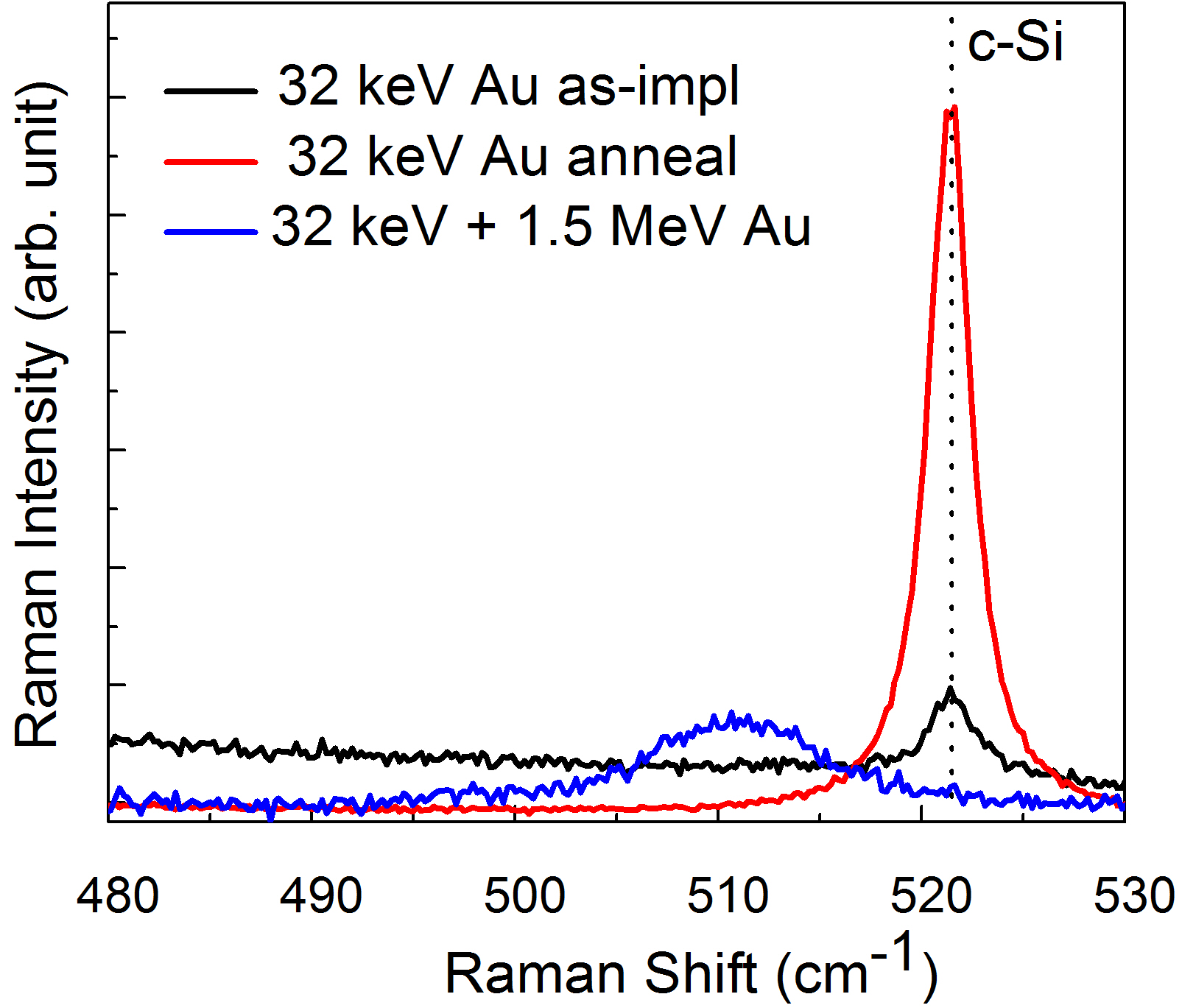}
\caption{Raman spectra of singly (both as-implanted and annealed at 500 C) and doubly Au implanted samples. The dotted line indicate Raman peak position of a c-Si.}
\label{raman1} 
\end{center}
\end{figure}

In case of crystalline Si (c-Si), the first-order Raman scattering probes the optical phonon frequency at the $\Gamma$-point in the Brillouin-zone due to the $q$ = 0 selection rule. 
This leaves a Raman active mode at 521 cm$^{-1}$, which gives a single line with a natural line-width of 3.5 cm$^{-1}$. At nano-scale, this selection rule is relaxed due to constraints imposed by uncertainty principle . There is a softening and broadening of the first-order phonon mode resulting in a shift of the Raman line towards lower wave-number together with a broadening in the line-shape. Compared to this, in a-Si, due to loss in long range order, the $q$-selection rule does not apply. As a result, all the phonons are optically allowed and the Raman scattering results in a broad hump at 480 cm$^{-1}$ \cite{ritcher}. It is also very important to mention that structural damage, strain, alloying etc can also result in a downshift and broadening in Raman spectrum and it is important to identify the exact effect. In Figure \ref{raman1}, Raman spectra of 32 keV Au implanted (single implantation) samples are shown for both before and after annealing at 500 $^o$C. One can see, there is a Raman peak at 521 cm$^{-1}$, with a rising tail towards lower wave-number which indicates the presence of amorphous phase in the system. After annealing, this rising tail disappears completely and the intensity rises by a factor of six. This shows that annealing at this temperature recrystallizes the amorphous region in the Si matrix. In addition, the absence of any structure below 520 cm$^{-1}$ also indicates the absence of Si NCs in the matrix. It is important to note that the damaged/amorphised layer thickness is only about 30 nm which is smaller than the optical penetration depth of the exciting laser radiation ( about 100 nm in a-Si). The observed 521 cm$^{-1}$ line from the single implanted sample, as seen in the as-implanted sample,  is therefore due to the c-Si matrix lying beneath the top amorphised layer. 

A subsequent 1.5 MeV Au irradiation with a fluence of $1 \times 10^{15}$ cm$^{-2}$ on the 
low energy as-implanted sample shows a  shift in Raman line towards the lower wave-number, $\sim$ 510 cm$^{-1}$. This is also included in the figure for comparison. However, there is 
no significant change in its relative intensity as compared to single 32 keV Au implantation.  On the other hand, in case of double implantation sample, the range of 1.5 MeV Au in Si is $\sim$ 400 nm which is much greater than the optical penetration depth ($\sim$ 100 nm) of exciting laser. Thus, here, the observed Raman signal is from the implanted region only.

\begin{figure}[h]
\begin{center}
\includegraphics[height=8.0cm]{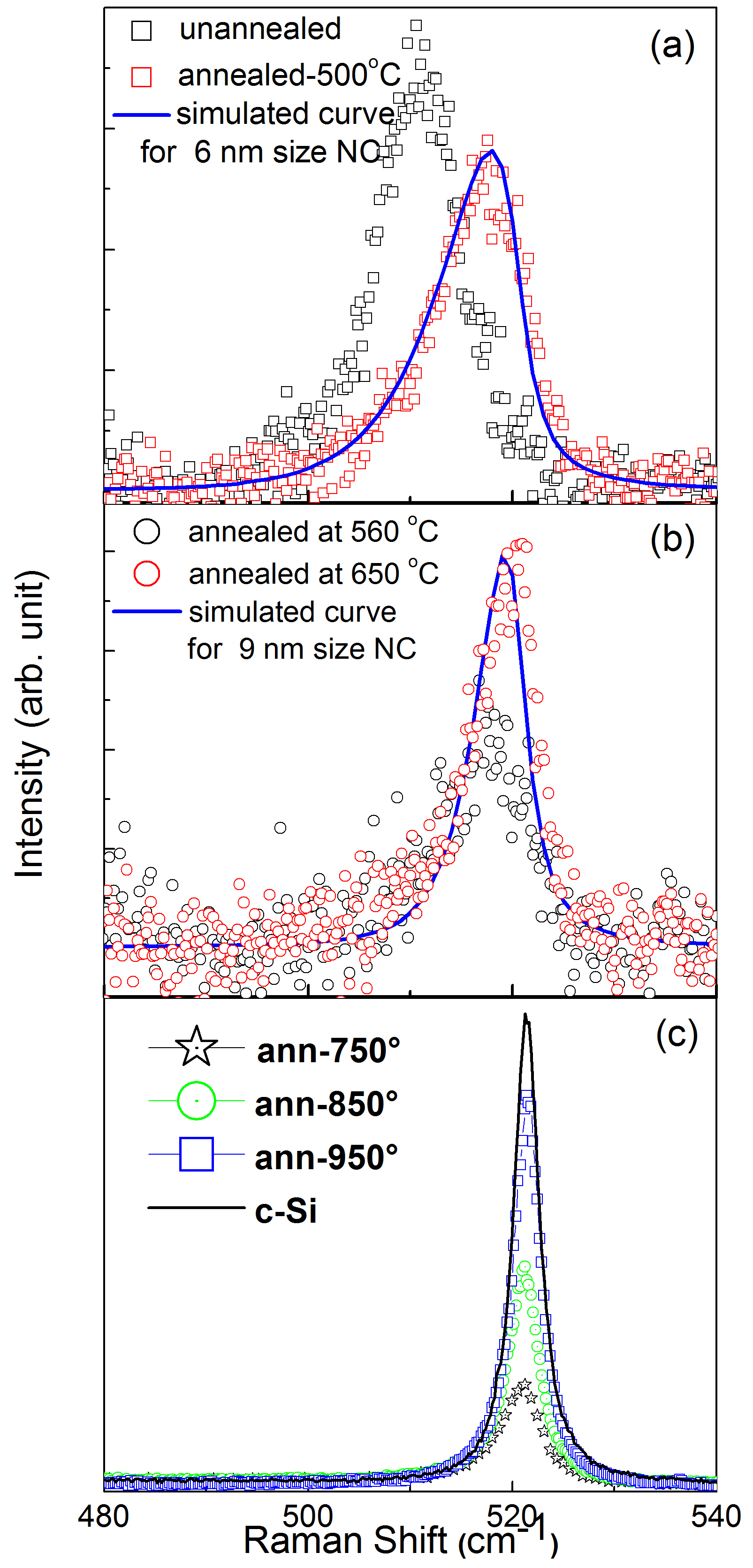}
\caption{Background subtracted Raman scattering spectra corresponding to (a) as-implanted and 500$^o$ C annealed sample, (b) sample annealed at 550, 650 C and (c) sample annealed at 750, 850 and 950 C. A spectrum of c-Si is also included for comparison.}
\label{raman} 
\end{center}
\end{figure}

The shift of the Raman peak from 521 cm$^{-1}$ to 510 cm$^{-1}$ in case of the doubly ion implanted sample can be either due to quantum confinement \cite{apl73-1568,pla205-117} or due to presence of strain induced as a result of ion-implantation. Quantum confinement effect is very unlikely to be the reason for the observed red-shift because the observed Raman line-shape is symmetric whereas the red-shift induced due to confinement of phonons are always accompanied by an asymmetry \cite{nrl3-105,jap101-64315}. On the other hand the
line-shape is broad as expected for NCs. Therefore, it is likely that the observed red-shifted symmetric Raman line is due to both strain and confinement effects \cite{mrs-proceeding}. In a previous similar work, using X-ray Photoelectron spectroscopy, we have shown that in the top a-Si layer, regions with complex Au-Si alloy are formed \cite{nanoGS}.  This has been confirmed from a binding energy shift of 1.2 eV and an extra broadening of the Au 4f spectrum in the implanted sample as compared elemental Au \cite{nanoGS}. This indicates the presence of Au-Si complexes in the Si NCs formed due to MeV ion implantation. This can introduce a tensile strain in the Si NCs resulting in a down-shift of Raman peak to 510 cm$^{-1}$. It is important to mention here that Au-Si bond does not have Raman active mode \cite{jjap34-5520}, due to which it is not possible to any signature of Au-Si formation in Raman spectrum. 

A clear analysis of quantum confinement effect on Si NCs  and the effect of annealing will be presented now. As shown earlier, the Raman line corresponding to doubly Au implanted sample, is seen to ride over a large background coming from the damaged layer. In order to see the Raman spectrum clearly it is necessary to subtract the background. Such data are shown in Fig. 2 (a)-(b) after subtraction of a linear background (between 480 and 540 cm$^{-1}$).  Raman spectra corresponding to samples annealed at 750$^o$ C or above do not required any background subtraction. This is due to the fact that annealing at higher temperature reduces the damage present in the system. After annealing at 500$^o$ C, the Raman peak shifted towards 521 cm$^{-1}$ without much change in line-width. As shown below the observed annealing dependent blue-shift of the Raman peak can be explained on the basis of a phonon confinement model (PCM) \cite{ritcher,campbell}. In this model, the first order Raman spectrum, in terms of intensity as a function of frequency is given by 

\begin{equation}
I(\omega)~=~\displaystyle\int^1_0\frac{exp(-q^2L^2/4a^2)}{[\omega~-~\omega(q)]^2~+~(\Gamma_o/2)^2} d^3q,
\end{equation}

where $q$ is expressed in units of $2\pi/a$, $a$ being the lattice constant, 0.543 nm. The parameter $L$ stands for the average size (diameter) of the NCs. $\Gamma_o$ being the 
linewidth of the Si TO phonon in bulk c-Si ($\sim$ 3.6 cm$^{-1}$). Following Tubino 
{\it et al.} \cite{tubino}, the dispersion $\omega$(q) of the TO phonon in a spherical Si NC
can be taken as 

\begin{equation}
\omega^2(q)~=~A~+~B~cos(\pi q/2),
\end{equation}

where $A~=~1.714 \times 10^{5}$ cm$^{-2}$ and $B~=~1.000 \times 10^5$ cm$^{-2}$. 
 
Using Eqns. 1 and 2, one can estimate the Raman scattered intensity spectrum for Si NCs of different diameters through a variation of $L$. From the above simulation, we observed that decrease in NC size results in both a broadening and an asymmetry in the Raman peak. In addition, all the  simulated curves of NCs at different sizes are found to be rising from 521 cm$^{-1}$ only. 

Annealing of the doubly as-implanted sample at different temperature is expected to result in the growth of NC in addition to removing ion-induced defects \cite{alford} in Si . There are reports on Si NC formation in SiO$_2$ using Si ion implantation at high fluence ($\sim ~ 10^{17}$ cm$^{-2}$ ) followed by high temperature annealing ($\geq$ 1050$^o$ C) \cite{apl76-351,jap83-6018,jap88-3954,sse45-1487}.  Present study reveals that ion implantation followed by annealing at a temperature as low as 500$^o$C can also lead to stable and strain-free Si NC formation. This is shown in figure \ref{raman}(a). One can clearly see a shift of the Raman line towards 521 cm$^{-1}$ indicating strain removal.
\begin{figure}
\begin{center}
\includegraphics[height=6.0cm]{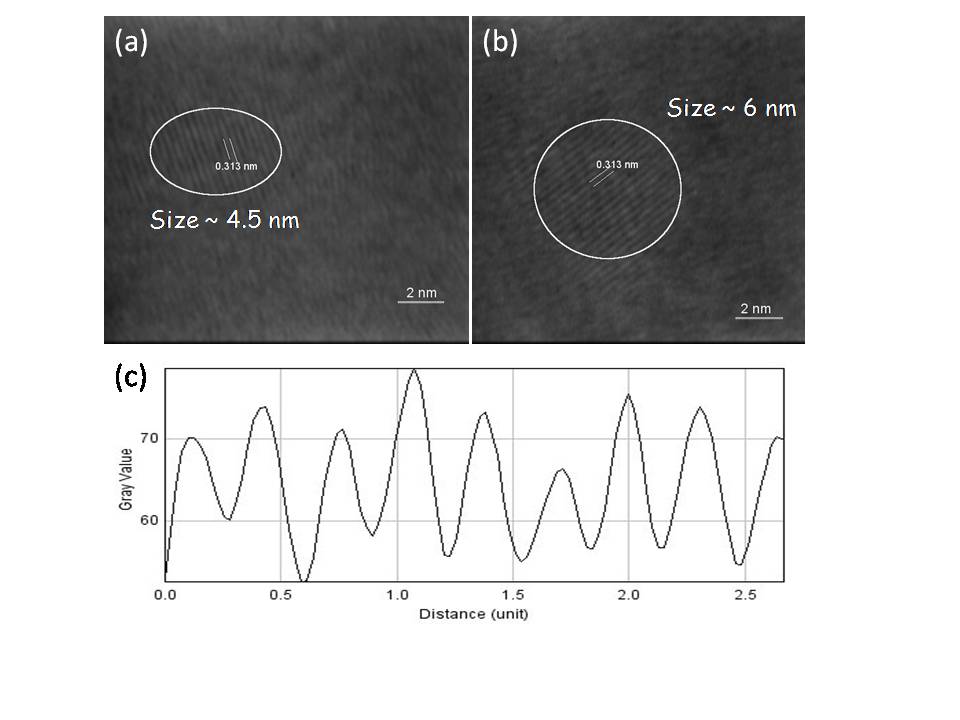}
\caption{A high resolution cross sectional TEM image taken on the sample.}
\label{tem} 
\end{center}
\end{figure}
The corresponding peak structure could be explained nicely using the 
PCM model for a given size. This is clearly shown in Fig. 2 (a) where the results following the annealing at 500$^o$ C are seen to exactly match that obtained for a NC size of 6 nm. Such spectra of few more samples, annealed at different temperatures 550$^o$, 650$^o$, 750$^o$, 850$^o$ and 950$^o$ C are shown in Fig \ref{raman} (b) and (c). Raman spectrum of a virgin c-Si sample is also included in Fig. \ref{raman}(c) for comparison. Annealing at 650$^o$ C is seen to result in NCs of average size 9 nm. Annealing at 750$^o$ C and above results in a stronger and narrower Raman peak appearing almost exactly at 521 cm$^{-1}$. This means that phonon confinement is very weak and the produced NCs size is much greater than 10 nm. Almost complete crystallization is seen to be achieved in case of 950$^o$ C annealed sample, as the Raman intensity and the width are both comparable to that of c-Si.  
%From these observations, we believe that with increasing annealing temperature NC/grain size %increases which lead to recrystallization of the Si matrix. 
Two cross-sectional TEM image corresponding to a sample annealed at 500$^o$ C are shown in Fig. \ref{tem}(a) and (b). The figures show NCs of size nearly 5-6 nm present in an amorphised medium. The separation between the planes, as shown in terms of grey values corresponding tocrystalline planes, is found to be 0.313 nm. This corresponds to  (111) planes of Si. 

\begin{figure}[h]
\begin{center}
\includegraphics[height=6.0cm]{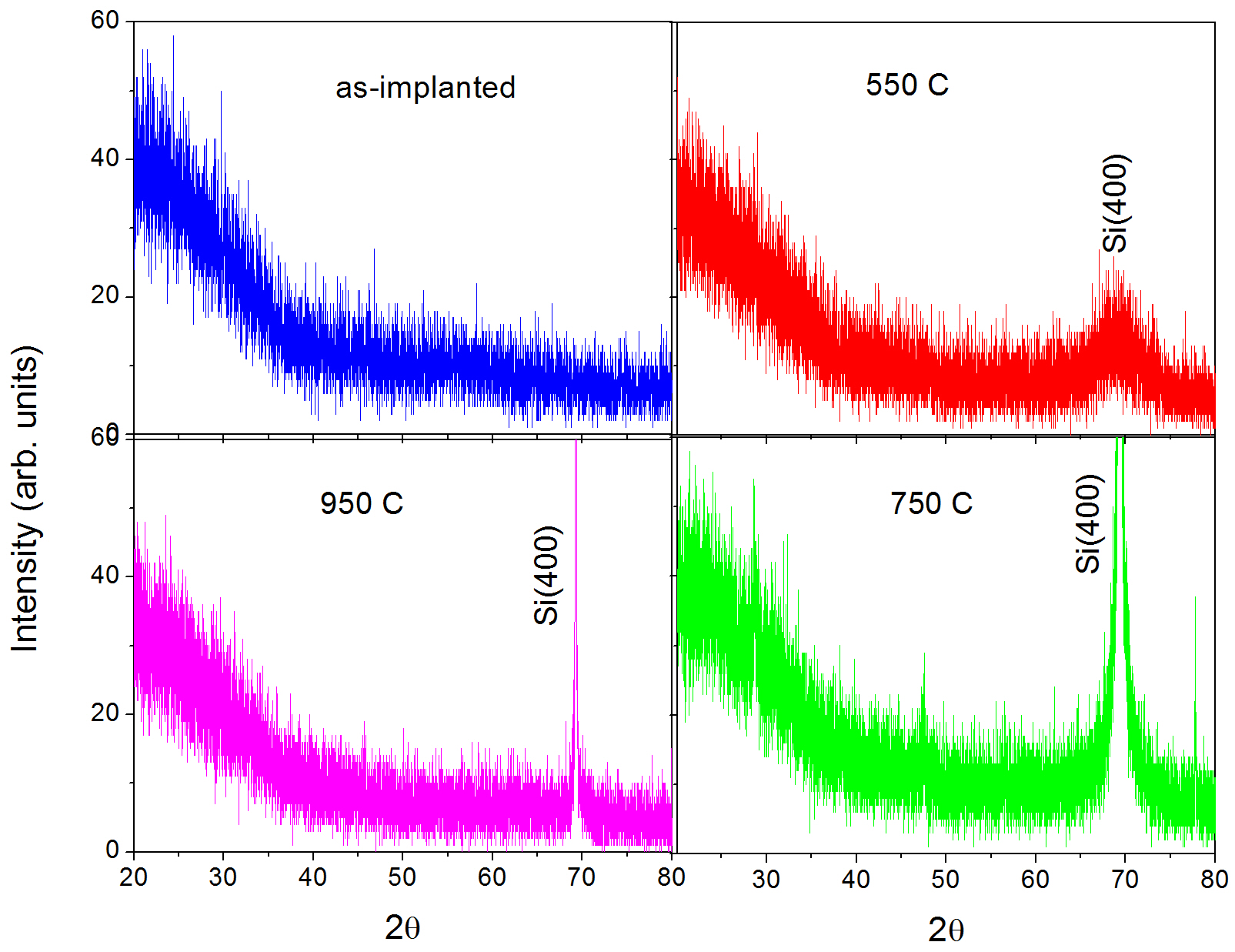}
\caption{ XRD results of doubly Au implanted samples before 
(top left) and after annealing at 550$^o$ C (top right), 
750$^o$ C (bottom right) and 950$^o$ C (bottom left).}
\label{xrd} 
\end{center}
\end{figure}

As has been mentioned earlier,  we have carried out XRD measurements on these samples. XRD results go in line with the above findings. Figure \ref{xrd} shows XRD spectra of as-implanted and annealed samples at different temperatures carried out at normal incidence. In case of an as-implanted sample, we have not found any structure. We believe, at this stage, the concentration of NCs is very low which increases with annealing temperature. In addition, annealing resulted in the movement of NCs towards the surface. On 
the other hand, annealing at 550 $^o$C and higher temperature do show a hump like structures at 69.4$^o$ corresponding to Si(400) reflection, which indicate the presence of Si NC. The widhth of the observed hump is found to decrease with annealing temperature, indicating increase in NC size. 

\subsection{Rutherford Backscattering Results :}

\begin{figure}[h]
\begin{center}
\includegraphics[height=5.0cm]{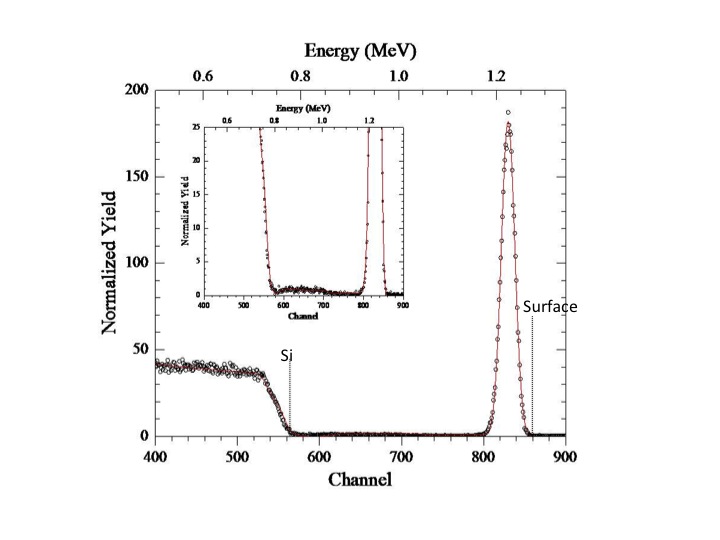}
\caption{RBS spectrum of as-implanted sample with simulated curve using 
RUMP}
\label{rbs-a} 
\end{center}
\end{figure}

The RBS spectrum of doubly Au implanted sample (before annealing) has been shown in Figure \ref{rbs-a}. A detailed simulation of the experimental data has been carried out using RUMP simulation package \cite{rump}; the result for the as-implanted case is included in the figure. The accuracy of the fit can be seen from the inset of the figure.  
Simulated data has been obtained using a three layer structure (layer1/layer2/layer3/Si-Substrate) where first layer is 22 nm Si, 2nd layer consist of 65 $\%$ Au and 35 $\%$ Si and third layer is Au tailing in Si with 0.9 $\%$ Au. As we know that the MeV implanted Au go much deeper (range of 1.5 MeV Au in Si $\sim$ 350 $\pm$ 70 nm), the observed near surface Au are due to low energy Au implantation. The simulation result agrees quite well with SRIM results (range of 32 keV Au in Si $\sim$ 22 $\pm$ 5 nm ). It is important to mention here that RBS analysis does not give any information regarding chemical composition/stoichiometry. We believe that the Au dominated region consists of pure Au particles together with Au-Si phases dispersed in the Si matrix, which also contains Si in amorphous and NC forms.  If Si NCs are present in Au dominated layer 2, it is likely that Au or Au-Si complex structures are present in these NCs. This induces tensile strain in the NCs which have been observed in Raman data as well. This reveals that the Si NCs are formed below the 20 nm from the surface. After annealing at 500 $^o$C or higher, these Au-Si structures get restructured and the Si NCs get freed from strain. Hence, true confinement type asymmetry is observed in Figure \ref{raman}.

\begin{figure}[h]
\begin{center}
\includegraphics[height=6.0cm]{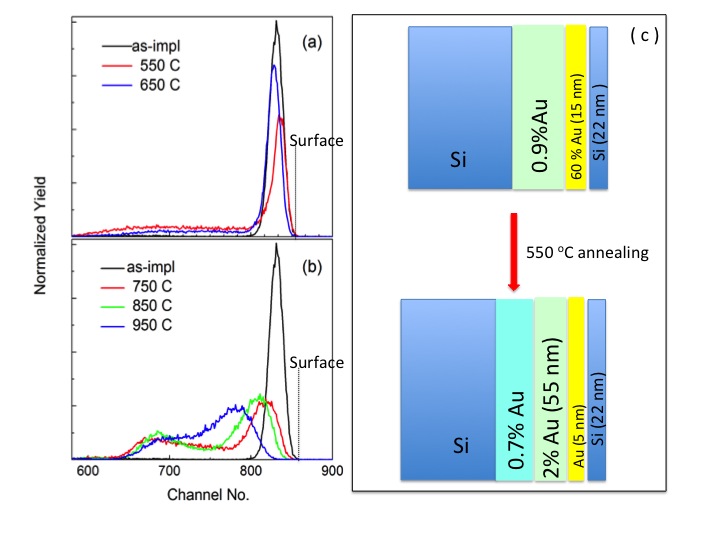}
\caption{(a) and (b) RBS spectra showing Au profile of doubly Au implanted 
samples for both as-implanted and annealed samples at different 
temperatures. (c) Schematic diagram explaining the annealing kinetics in the system.}
\label{rbs} 
\end{center}
\end{figure}
	
The RBS spectra corresponding to annealed samples at different temperatures i.e 550, 650, and 750, 850 and 950 $^o$C has been shown in Figure \ref{rbs} (a) and (b) respectively. 
Our previous work on single keV implantation followed by annealing reveals inward diffusion of Au in Si \cite{nimbGS}. Followed by that, in the present study, we expect the keV implanted Au diffuses towards the Au/Si interface whereas the MeV implanted Au out-diffuses to the surface resulting in a bi-modal distribution of Au in Si matrix. This bi-modal distribution of Au is clearly seen at the annealing temperature of 850 $^o$C (Figure \ref{rbs} (b)). The reason behind such a Au split is mainly due to the presence of dislocations/extended defects (because of implantation), which provides a means for rapid pipe diffusion \cite{alford}. A detailed RUMP analysis of the annealed samples reveals that Au-Si layer thickness decreases and Au layer rises as we increase the annealing temperature. 
Simulation for the annealed sample at 550 $^o$C, shows that the pre annealed layered structure gets rearranged (due to annealing) and a 5 nm layer consisting of Au nano/micro-particles \cite{nimbGS} is present between layer 1 and layer 2. A schematic diagram explaining the above has been shown in Fig. \ref{rbs} (c). Annealing at higher temperature than the eutectic temperature (363 $^o$C) of Au-Si system, forms alloy of Au and Si, taking more and more Si atoms from the matrix. During cooling, Si atoms precipitate from Au-Si \cite{apl92-103105}, separating Au and Si. In the present case, annealing resulted in removal of Au from Si NCs, leading to a strain-less NCs. Few of Au atoms diffuse towards the surface and forms Au NCs, which is seen in SEM \cite{ftcGS}. This supports our arguments that with increasing annealing temperature Au-Si structures present in NCs get restructured and strain free NCs are formed in annealed samples.

\section{Conclusions}

In summary, the dependence of the size of Si NCs embedded in bulk Si matrix on the annealing temperature was investigated by Raman Scattering and XRD. A two-stage Au implantation and irradiation technique followed by low temperature thermal annealing has been used to successfully synthesize Si NCs embedded in the top amorphous Si layer. %Initially Si samples were implanted with 32 keV Au ions to amorphize the top surface layer. A subsequent 1.5 MeV Au irradiation has been found to result in formation of strained Si NCs in the top a-Si layer through ion beam induced localized crystallization. The observed tensile strain is seen to be removed following an annealing at 500$^o$ C for an hour. 
Annealing of samples at 500 $^o$C and higher lead to growth of  NC produced in the bulk Si matrix. And almost complete recrystallization has been seen following an annealing to a temperature 950$^o$ C. 
%All the Au are concentrated at some depth in a single layer in case of subsequent annealing which may proved to be good !!!!
Most important finding here is Raman scattering results indicate a very narrow size distribution of the NCs formed in the matrix. This technique which results in production of Si NCs with $\delta$-function like size distribution is expected to be very useful for future applications due to its uniformity in size.

\section{Acknowledgements}
One of the authors (GS) would like to thank the technician staffs of Ion Beam Laboratory of Institute of Physics Bhuabneswar, for their helps during implantation and RBS measurements. The author would also like to acknowledge DST, Govt. of India, for the funding under DST Fast Track Scheme for Young Scientists, project No. SR/FTP/PS-007/2012.

\end{document}